# Performing energy modelling exercises in a transparent way - the issue of data quality in power plant databases

*Fabian Gotzens[1][a], Heidi Heinrichs[a], Jonas Hörsch[b,c], Fabian Hofmann[b]*

[a]Institute of Energy and Climate Research (IEK), Systems Analysis and Technology Evaluation (IEK-STE), Forschungszentrum Jülich, Wilhelm-Johnen-Straße, 52425 Jülich, Germany
[b]Frankfurt Institute for Advanced Studies (FIAS), Ruth-Moufang-Straße 1, 60438 Frankfurt am Main, Germany
[c]Institute for Automation and Applied Informatics (IAI), Karlsruhe Institute of Technology, 76344 Eggenstein-Leopoldshafen, Germany

**Abstract**

In energy modelling, open data and open source code can help enhance traceability and reproducibility of model exercises which contribute to facilitate controversial debates and improve policy advice. While the availability of open power plant databases increased in recent years, they often differ considerably from each other and their data quality has not been systematically compared to proprietary sources yet. Here, we introduce the python-based 'powerplantmatching' (PPM), an open source toolset for cleaning, standardizing and combining multiple power plant databases. We apply it once only with open databases and once with an additional proprietary database in order to discuss and elaborate the issue of data quality, by analysing capacities, countries, fuel types, geographic coordinates and commissioning years for conventional power plants. We find that a derived dataset purely based on open data is not yet on a par with one in which a proprietary database has been added to the matching, even though the statistical values for capacity matched to a large degree with both datasets. When commissioning years are needed for modelling purposes in the final dataset, the proprietary database helps crucially to increase the quality of the derived dataset.

**Keywords:** open data; power plant data; Europe; power system model; energy system analysis

---

[1] Corresponding author: e-mail f-gotzens@fz-juelich.de, phone +49(2461)619834, fax +49(2461)612540

# 1 Introduction

In energy modelling the question of traceability and reproducibility of model exercises has been heavily debated in recent years. It is emphasised how important both open data and open source code are in this context to allow for controversial debates of model outcomes which often serve as policy advice (cf. [1], [2], [3], [4]). One of the big challenges for policy makers today is to manage the transition of energy systems towards sustainability. As sustainability is not only limited to technically or economically feasible solutions, it requires social feasibility in aspects like justice or acceptance as well. Therefore, it is crucial for a successful energy transition to discuss different competing pathways with varying benefits for different groups in society openly with all stakeholders. This represents a highly complex task suited to be addressed by modelling exercises.

In this sense, energy system modellers face the challenge to provide at least all relevant assumptions in a traceable and transparent way (cf. [5]). In this context we want to point out the difference between transparency and openness. While transparency requires a study to be traceable in terms of all assumptions made, openness even calls for open access to the applied code and used data. However, none of them guarantees reproducibility as transparency does not always necessitate 'completeness' of information (data and code) and openness could lead to a confusing quantity of information hindering application or understanding. In this respect, one should not take openness or transparency directly as reproducibility, in fact they are rather preconditions for reproducibility, but do not guarantee it. Additionally, open data sources show different levels of data quality which again is often hard to judge objectively. Nonetheless, this does not call for avoiding openness or transparency, but rather intends to raise awareness of related dangers (for further information on challenges related to openness cf. [6]).

Up to now, no systematic analysis exists as to whether a certain source of open data has an inherent lack in data quality compared with proprietary data sources or to what extent proprietary data sources might outperform currently available open data sources. However, one needs to be aware that this question can only be answered for specific cases. Here, we discuss this issue of data quality for power plant databases which are used as key input to various kinds of energy modelling exercises. We chose this case due to the already extensive availability of open data for conventional[2] power plants (e.g. [7], [8], [9], [10], [11], [12]). Nonetheless, the same issue of data quality applies to all required input data describing the current, and with certain reservations regarding its assumptions also the future, framework of the system under consideration (e.g. electricity demand, grid constraints, $CO_2$ caps, fuel prices, etc.). Therefore, we focus here on challenges related to data mining for modelling needs while we describe applications of modelling exercises using the derived datasets in [13, 14].

Here, we present an approach on how to derive and check transparent and open power plant fleets for each European country. We introduce *powerplantmatching* (PPM) [15], a tool for merging power plant databases into a final dataset and apply it to evaluate several combinations of open and proprietary databases of power plants.[3] This allows us to evaluate differences in the used input data and how those databases complement one another. Overall we aim at identifying if a dataset derived purely from open data can compete with a dataset derived from open and proprietary databases by comparison with national capacity statistics.

## 2   Methodology

To derive a consistent power plant fleet for each European country we developed powerplantmatching (PPM) that is a toolset for cleaning, standardizing and combining multiple power plant databases [15]. At first, we give an overview of the databases which were used as inputs in PPM in subsection 2.1 and then we present briefly how PPM works in subsection 2.2. We show how we rescale between gross and net capacities to allow for comparability in subsection 2.3 and how we deal with wind and solar power units in subsection 2.4. The methodology section closes with a plausibility check as a proof of concept which is described in subsection 2.5.

---

[2] We excluded wind and solar units from our analysis, cf. subsection 2.4.

[3] In order to avoid confusion, the following wording has been agreed upon: For the raw collections of input data we use "database", while "dataset" is used for the processed versions including the matching result and "fleet" means the power plants claimed for one single country. While "Power plant" refers to an entire generating station at one location, "unit" and "block" are used interchangeably and refer to a fraction of a power plant.

*2.1 Short overview of used databases*

For PPMs application six databases that are openly available[4] and one proprietary database have been used (cf. Table 1) which all differ in their magnitude both in number of units and represented capacities but also in geographical scope, level of detail and their definitions (e.g. fuel types). All of these databases have been filtered such that they only contain units within the geographical scope of interest, here we chose EU28 + Switzerland + Norway - Cyprus - Malta. However, the end user is free to choose his individual geographical coverage in the config file of PPM. The capacities in the different databases range from small kW-scale units up Europe's biggest plant, Bełchatów (5.42 GW) in Poland. An overview of the massive number of records being contained in the different databases, especially those in CARMA and WEPP, and their distribution in terms of capacities and fuel-types the is shown in Figure 1.

---

[4] Meaning that they are either published under an open source license or are freely available for download.

**Table 1: Applied power plant databases that cover the EU-28 + Switzerland + Norway - Cyprus - Malta [7], [8], [9], [10], [11], [12] and [15]**

| Database Supplier | Abbreviation | Type | No. units | GW |
|---|---|---|---|---|
| Carbon Monitoring for Action | CARMA | gross | 50,570 | 4,931.96 |
| European Network of Transmission System Operators for Electricity | ENTSOE | net | 4,384 | 851.14 |
| DOE Energy Storage Exchange (only pumped storages) | ESE | net | 850 | 153.72 |
| Global Energy Observatory | GEO | gross | 1,314 | 692.02 |
| Open Power System Data (Conventional Power Plants) | OPSD[5] | net | 6,768 | 571.08 |
| World Electric Power Plants Database | WEPP | gross | 63,398 | 1,848.83 |
| World Resources Institute | WRI | unknown | 2,867 | 360.898 |

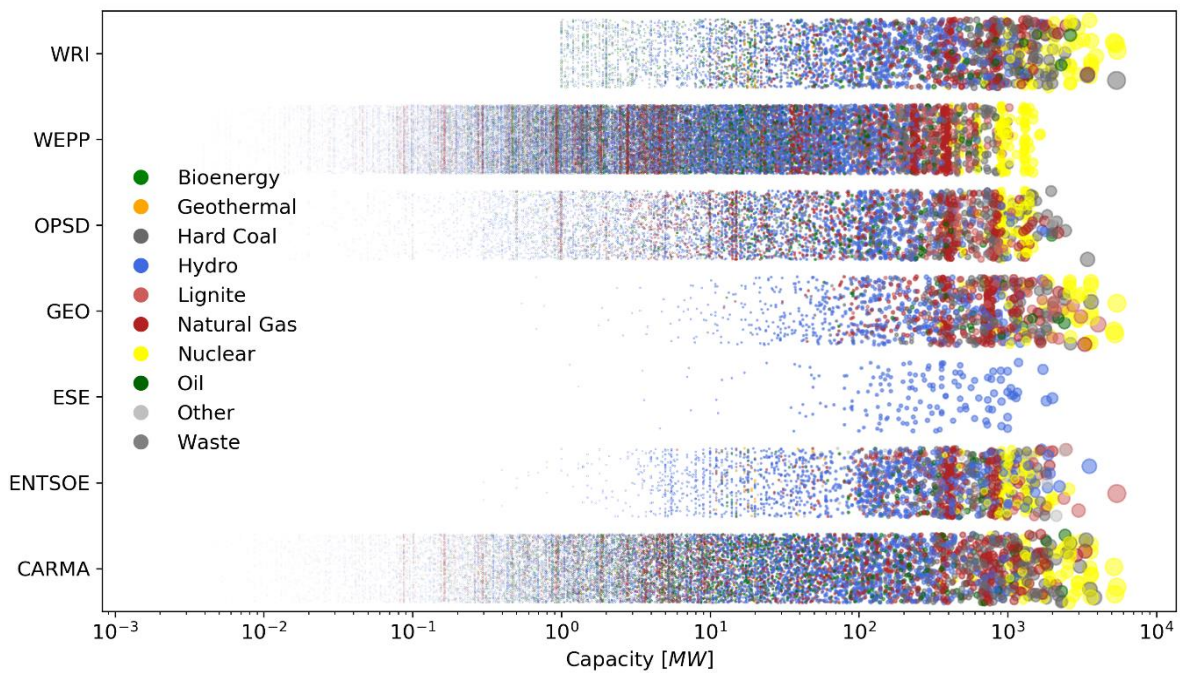

**Figure 1: Overview of the different input databases. Each marker represents one record, is scaled based on its capacity and colored based on its fuel type.**

The proprietary *World Electric Power Plants Database* (WEPP) is an example of one of the most widespread power plant databases used by academics, NGOs and businesses [16, 17]. It is the only database that has been acquired commercially beforehand in the version from

---

[5] Note that the capacity statistics from *ENTSO-E: Scenario Outlook and Adequacy Forecast (SO&AF)* as reported by OPSD are not to be confused with the conventional power plant database provided by OPSD.

September 2016. Even though it is allegedly updated quarterly according to the vendor, it contains units that one does not find any information about elsewhere, neither in the other databases nor in the internet through manual research (e.g. 'Aachen Works 1'). The fact that it also does not contain geographic coordinates of the plants adds doubt as to whether it always reflects real existing units at its release date. In contrast, *Open Power System Data* (OPSD) is "a free-of-charge data platform dedicated to electricity system researchers" providing "data on installed generation capacity by country/technology, individual power plants (both conventional and renewables-based), and time series data" in the form of individual data packages [18]. In the following, we refer to the conventional power plants data package as OPSD, which is used as input for PPM. Since the *DOE Energy Storage Exchange* (ESE) [11] is a database of storage units, we filtered it such that in our case it only contains pumped storages, as smaller storage systems like batteries are not contained in the other database and could therefore not be successfully matched. The database provided by the *World Resources Institute* (WRI) obviously does not contain any units below MW-scale. PPM obtains the ENTSOE[6] database directly through the application programming interface (API) of the *ENTSO-E Transparency Platform* [10] to keep it up to date. Nevertheless, there are examples of power plant owners reporting less than the full net generating capacity to the *European Network of Transmission System Operators for Electricity* (ENTSO-E); possibly, to reserve a certain part of their plant's capacity for spinning and control reserve. For instance, Germany's hard coal and natural gas power plant 'Gersteinwerk' has a total nameplate capacity of 2,372 MW (2,040 MW excluding gas turbines) [19], while ENTSOE only reports 2,003 MW of total installed capacity. All of the aforementioned differences add to the challenge of matching. For example, power plants which were historically fuelled with coal but have been retrofitted to natural gas at some point, might be given with coal in one database and with natural gas in another.

## 2.2 Brief introduction to PPM

PPM has been implemented in Python and is available under a GNU/GPLv3 licence. In order to understand how the underlying method of PPM works, we describe the main steps and program modules in the following. Basically the PPM method can be broken down into four steps: (1) standardization of the terms and tabular structure for each database, (2) aggregation of units into power plants, (3) linkage of the aggregated power plant lists, and (4) reduction of the connected claims (cf. Figure 2).

---

[6] We intentionally renounced the hyphen here, enabling us to distinguish clearly between the organization (ENTSO-E) and the obtained database used as input for PPM (ENTSOE).

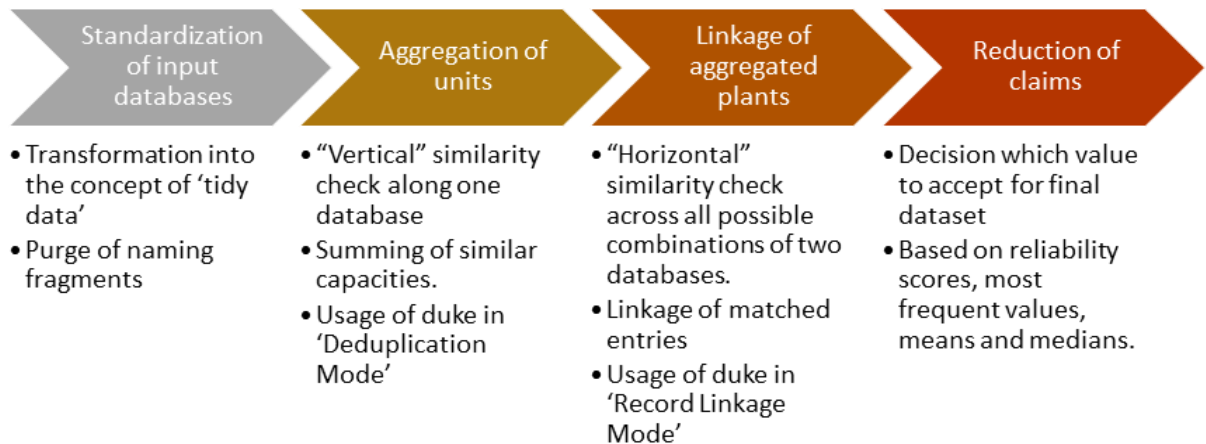

**Figure 2: Flow chart of the main steps in PPM**

The first step comprises the *data* module of the PPM package, which compiles explicit mappings to translate the different terms and structures of each database into a previously defined structure (cf. Table 2) with respect to the concept of 'tidy data' [20]. For example, a mapping of the fuel types of the WEPP can be found here [21]. In addition, databases exposing gross power capacities are rescaled. The used scaling factors are described in subsection 2.3 more in detail.

**Table 2: Standardized data structure**

| Column heading | Description | Example(s) |
|---|---|---|
| Name | Name for the unit/plant | "Bouchain 7", "Centrale Maasvlakte" |
| Fueltype | General fuel type | "Wind", "Solar", "Natural Gas" |
| Technology | Further specification | "Onshore", "PV", "CCGT" |
| Set | Indicator if CHP used | "PP" or "CHP" |
| Country | Short country name | "France", "Germany", "Latvia", ... |
| Capacity | Net/gross installed capacity | 645.0 |
| YearCommissioned | Year when unit came online | 1995 |
| lat | Geographical latitude | 51.96262 |
| lon | Geographical longitude | 4.025152 |
| File | File of origin | RTE |
| projectID | ID in original file | OEU123 |

Most power plant databases like OPSD, ENTSOE, ESE and WEPP report individual power plant units, however, aggregated power plants are commonly integrated in models on system scale, so an aggregation is needed (step 2). This aggregation step takes place before different databases can be compared to each other and is based on the approximate probability that any two units belong to the same power plant. This is computed by weighting the similarity between name, fuel type and geographic location with a naive Bayesian classification scheme implemented in the java application Duke [22]. Groups of units with pair-wise similarities above a high threshold (98.5%) are collected as power plants. The capacities of the power plant units belonging together are summed and the most frequently occurring name is kept while the geo-coordinates are being averaged.

The core of the PPM tool is the third step, which links the separate power plant datasets. The same comparison scheme based on Duke with slightly different weights is used to determine similar power plants for every pair of datasets. These individual links from dataset to dataset are iteratively joined to chains connecting as many datasets as possible. Each chain links several - sometimes conflicting - claims about the same power plant. These claims are then reduced according to a predefined reliability score (cf. Table 5). Sources which were both updated recently and checked manually (OPSD) get the highest score, before sources that are only updated regularly (ESE, ENTSOE and WEPP), before sources which have not been updated for a longer period of time (GEO, WRI, CARMA). The claim originating from the database with the highest score is then accepted for the final dataset. However, it can happen that claims with the highest but same score stem from two or more databases. In this case the acceptance is based on the most frequent name, fuel type, and technology in addition to the mean location in terms of latitude and longitude and the median capacity. In general, it is important to note that at least two input databases are needed for one record (=power plant) to occur in the final matched dataset.

For a slightly more detailed description of the specific algorithms used in PPM please refer to Section 2.2 in the companion paper [14].

## 2.3 Rescaling between Gross and Net Capacities

Since several databases provide gross capacities (e.g. WEPP and CARMA) whereas others (e.g. ENTSOE) provide net capacities, these values need to be standardized, i.e. rescaled to either of them, before a sensible matching can take place.

The OPSD database for Germany, which was put together manually by the OPSD modellers, is based on two different originating databases, one from the *Federal Network Agency* (BNetzA) and one from the *German Environment Agency* (UBA). While the former one provides unit sizes as net capacities, the latter provides them as gross capacities. This is of threefold advantage giving us the chance (a) to derive fuel- and technology-specific correction factors, (b) to check how PPM's vertical aggregation algorithm performs and (c) to evaluate the horizontal matching process of PPM (cf. subsection 2.4).

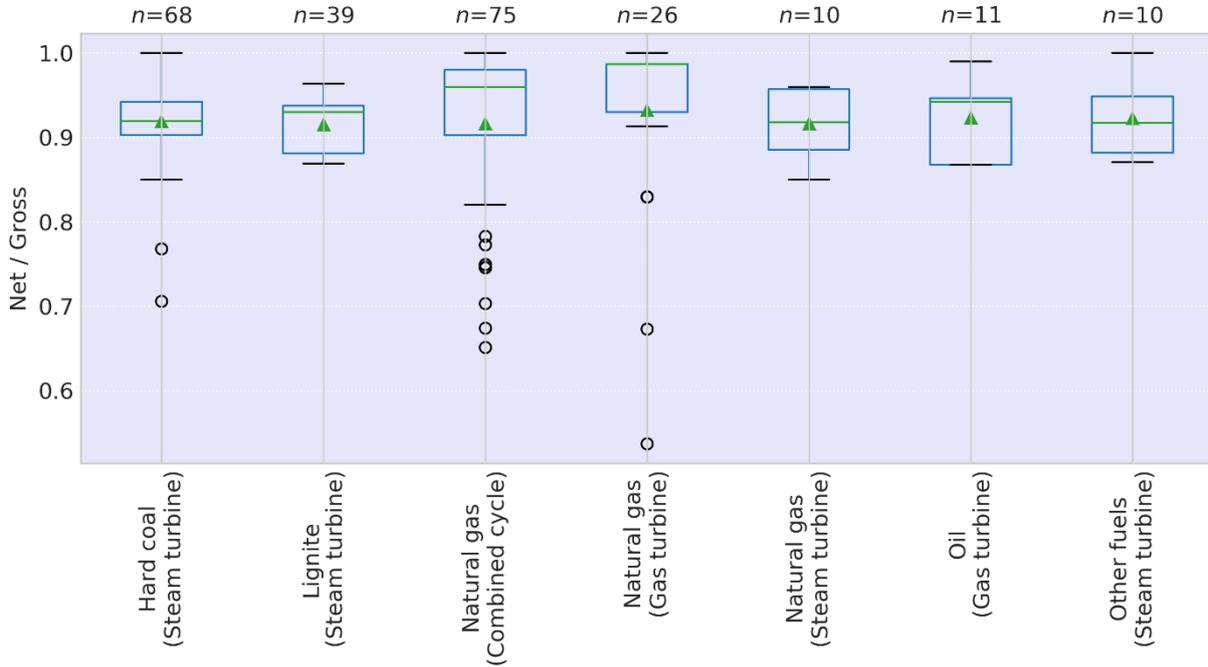

**Figure 3:** Boxplot of the ratios between net and gross capacities for different fuel-types and technologies. Triangles indicate means, green horizontal lines medians and circles outliers.

Figure 3 depicts the ratios between net and gross capacities for a combination of fuel-type and technology in the form of boxplots. We recognize that all medians and all but one means are above and close to 0.9 confirming the rule of thumb that the net capacity of power plants is usually about 90% of their gross capacity due to their internal consumption [23]. However, the diagram also displays a couple of outliers, which are defined as values that are either lower than the 1$^{st}$ quartile minus 1.5 IQR (inter quartile range) or higher than the 3$^{rd}$ quartile plus 1.5 IQR. Those have been investigated further manually and can be grouped into two categories:

(1) Outliers for (Natural Gas, Combined Cycle) stem from industrial plant owners, like chemical and automotive companies as well as refineries. Therefore, by nature their net capacities are smaller than their gross capacities of the boilers.

(2) The remaining outliers have been checked individually, i.e. their given net to their gross capacities have been compared to those given by owners of the units and seem to simply result from input errors. Hence, their rescaling factors are not representative of their combination of fuel type and technology.

After having confirmed that all groups of outliers can be explained, this demonstrated to a great extent why the averages are mostly lower than the medians. In order to represent these outliers in the rescaling process, we consequently decide to use the mean values as rescaling factor (based on the combination of fuel type and technology as given in Figure 3).

*2.4 Dealing with Wind and Solar Power Plants*

Since especially wind and solar units are comparatively small in terms of unit-wise capacity (ranging from very-low kW to low MW scale), but huge in terms of deployed numbers (e.g. more than 1.7 million single units only in Germany [24]), they would massively impede the matching process. Moreover, since most of the single solar panels and wind turbines do not have a specific name (except larger wind and solar parks), there is no data in the name column and, consequently, no input for the string comparator available. Therefore, all wind and solar units are being filtered as part of the data mending, enabling us to keep the entire process computationally manageable.

However, for a further usage of the derived power plant data in modelling approaches, PPM is able to concatenate given wind and solar units from the OPSD renewable data package [24] to the final dataset at the end of the matching process.

*2.5 Plausibility Check: UBA vs. BNetzA*

As mentioned above in subsection 2.3, the OPSD conventional power plant database for Germany has been assembled by the OPSD modellers by manually linking and merging the two source databases BNetzA and UBA. The 413 collected links are published as part of the package. Unfortunately, changes in the power plant list on identifiers or operating status in the source datasets since the last update on July 14, 2016 have invalidated all but 319 links, illustrating the need for a mostly automatic linking scheme. Aggregating the power plant units of BNETZA and UBA separately as described in Section 2.3 determines 176 power plants in BNETZA and 166 power plants in UBA connected by 181 manual links. The discrepancy derives from overlapping aggregation groups of power plant units identified by PPM, which mostly result from different fuel type specifications for the same block in BNETZA and UBA. PPM finds 153 correct links and no wrong link. Of the 28 missed links 21 are hidden by the incorrectly chosen aggregation groups.

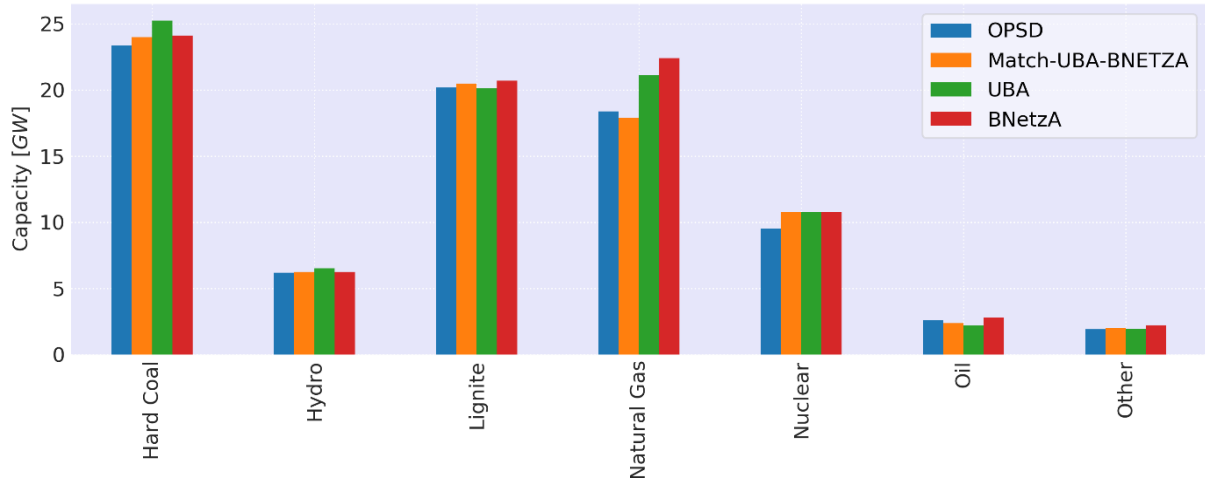

**Figure 4: Plausibility check of PPM results for Germany**

Figure 4 illustrates the capacities per fuel type for the PPM matching result (orange) of the UBA (green) and BNetzA (red) databases together with the comparison to the reference database from OPSD for Germany (blue). The comparison shows that the algorithm is able to reproduce OPSD's manual matching for the German fleet closely for almost all fuel types. Since block B of *Gundremmingen nuclear power plant* was already marked as 'shutdown' in OPSD in 2017, this explains the missing of 1,284 MW in the blue bar, so it matches even exactly for all remaining nuclear units. The lowest underestimation occurs for Natural Gas (-0,482 MW) and the highest overestimation for Hard Coal (+0.627 MW). The remaining differences result from different operating statuses of certain power plant blocks, since some of them are currently in an intermediate state between operation and shutdown, in the databases referred to as "temporary shutdown", "security reserve" or "special case", making it hard to distinguish the real operating plant size.

Altogether, we can state that our matching algorithm produces plausible results as it is able to reproduce the reference capacities (82.09 GW) to a very good extent (83.76 GW) with an absolute deviation of only about 2%.

## 3   Analysis

The analytical part of this paper is twofold: In the first part, we assess to what extent open databases cover Europe's installed capacities by comparing them among themselves and also to national capacity statistics. The latter are also reported by and taken from the OPSD-initiative through their *national generation capacity data package* [25] and must not be confused with their *conventional power plants data package* (cf. subsection 2.1). Even though the initiative gathered statistical data from many different sources, only the data taken from *ENTSO-E's Scenario Outlook and Adequacy Forecast SO&AF*) covered Europe completely in the base year 2016, therefore we used it here. The statistical values serve as a reference to a, yet unknown, reality since these statistical values contain uncertainties themselves and, of course, do not contain specific power plants. For all variants, the year 2016 represents the base year, as being the most recent year for which the statistics contain full historical information. In the second part, we check how much the proprietary WEPP database can contribute to the matching by adding it to the matching process.

*3.1  Part I: Assessing the coverage of open databases 'on their own'*

In this part, we show how PPM is applied to six open power plant databases (all but WEPP from Table 1) that went through the matching process.

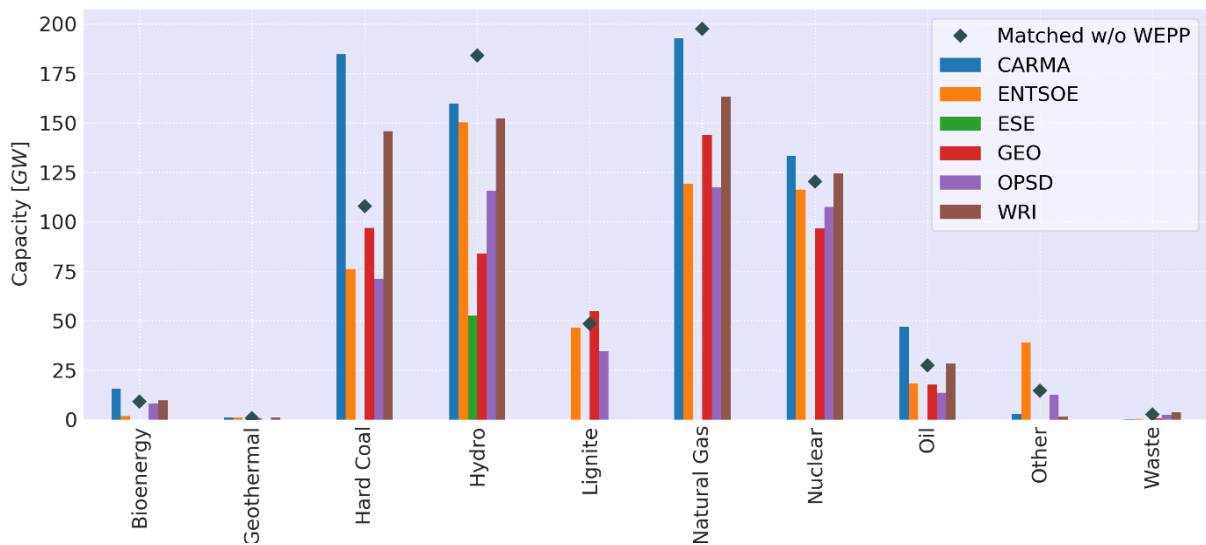

**Figure 5: Fuel type-specific capacities per input dataset as bars and the matching dataset as diamond marker**

Figure 5 depicts the fuel type-specific data for each input database as bar chart and displays the matching result with diamonds. The chart shows clearly that the initial databases differ vastly among each other and that CARMA contains the highest capacities for all fuel types, apart from lignite, in which it does not contain any plants. However, since CARMA contains more than double the capacity of hard coal than ENTSOE, this indicates that the database providers have not distinguished between hard coal and lignite. As previously stated in subsection 2.1, ESE only contains pumped storages, which are, of course, classified as hydro plants. Since for a positive match of one single power plant at least two databases are needed, in theory the matched dataset can contain only the maximum capacity of one database if two identical databases were fed into PPM in theory. However, since more than two databases are part of the matching process, the summed capacity in the matched dataset can, of course, be higher than the maximum of one of the input databases. In practice, the claimed capacity in the final dataset is always lower than the maximum of the input databases, except for hydro power.

The capacities of the resulting matched dataset are now grouped by fuel type and compared to statistical values, displayed in Figure 6. Unfortunately, the statistics report some fuel types in an aggregated form: "Bioenergy and other renewable fuels" were assigned to "bioenergy" and "Bioenergy and renewable waste" to "waste". All of the following aggregates, namely "Differently categorized fossil fuels", "Differently categorized renewable energy sources", "Mixed fossil fuels", "Other or unspecific energy sources" and "Tide, wave, and ocean" were assigned to "other". This leads to relatively high statistical values for "other" and "waste" and to very low values for "bioenergy". Still, for the non-aggregated fuel types obvious deviations occur, with the exception of nuclear plants.

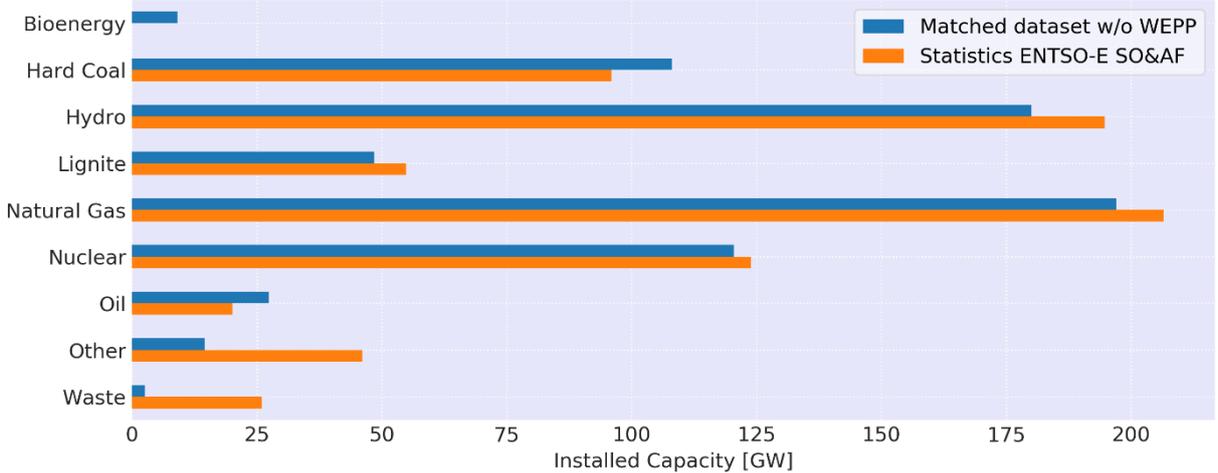

**Figure 6: Capacities of the matched dataset and statistical values by fuel type**

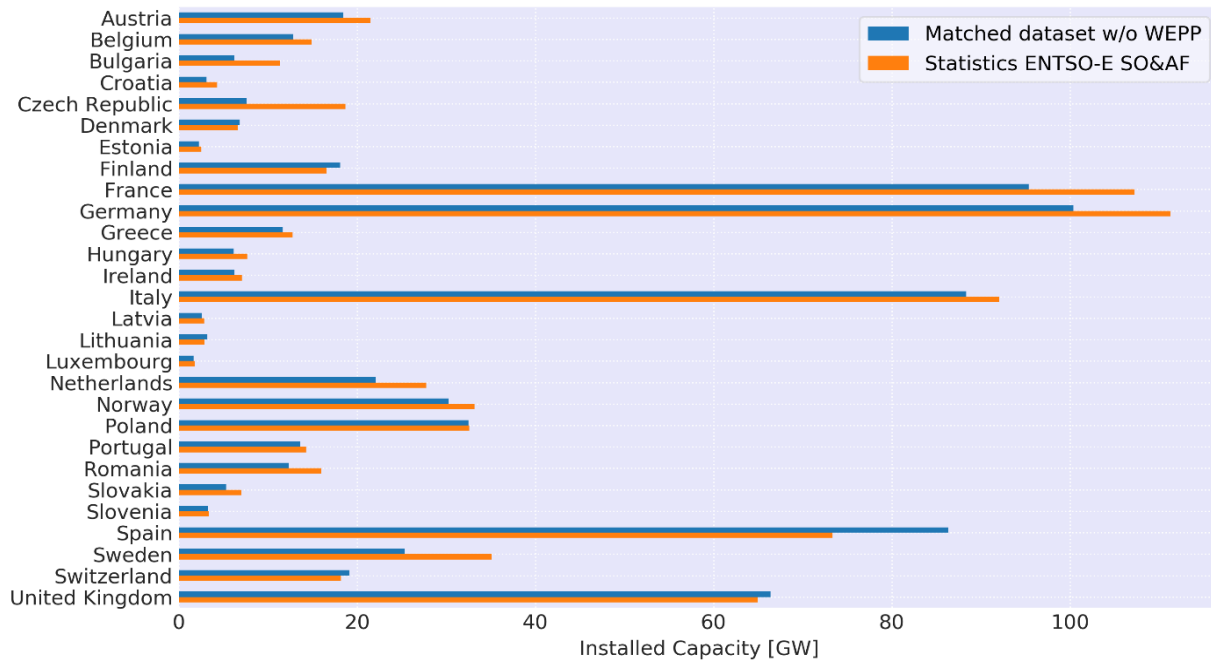

**Figure 7: Capacities of the matched dataset and statistical values by country**

In order to analyse these deviations from a different perspective, a plot showing the same data, but grouped by countries instead of fuel types, has been created (cf. Figure 7). It shows effectively that, in particular larger countries, namely France, Germany, and Sweden, are not yet covered very well, while e.g. Poland and Denmark match to a high extent, whereas some overestimation occurs in Spain. The discrepancy between the statistically reported total capacities (~768 GW) and those of the matching result (~708 GW) for the above-mentioned continental European countries can hardly be neglected even in view of inherent uncertainties. Therefore, we investigate in the following subsection, whether the missing capacity is provided by a proprietary dataset.

### 3.2 Part II: How much can WEPP contribute to open databases?

Up to here, PPM's application had only been focussed on the matching of freely available databases. Now, we extend this matching by adding WEPP as input database into the matching process. We define four objects of comparison, which form the basis of our analysis here. First, a matching result including the WEPP (i). Second, the matching result without the WEPP (ii), already known from part I (cf. subsection 3.1). Third and fourth, we take the WEPP only (iii) and, again, capacity statistics from ENTSO-E SO&AF (iv) for comparison into account. We chose these objects to identify to which extent the WEPP contributes to the matched database and to evaluate the differences to the WEPP and statistics alone. In doing so, we can test the level of suitability of open available power plant databases for modelling purposes. However, we must keep in mind that the WEPP has been acquired in the version as of 09/2016 (cf. subsection 2.1), so it might miss capacity additions and/or retirements of the fourth quarter of that year. Hence, we acknowledge that WEPP, just as capacity statistics, contains uncertainties itself and can, therefore, only be seen as an approximation of the hidden real-world capacity installations.

**Table 3: Cumulative installed capacities**

| No. | Object of Comparison | Records [-] | Capacity [GW] | Ratio to Statistics |
|---|---|---|---|---|
| i | Matched dataset w/ WEPP | 14,348 | 747.41 | 97.32% |
| ii | Matched dataset w/o WEPP | 6,014 | 707.65 | 92.14% |
| iii | WEPP only | 36,796 | 728.65 | 94.88% |
| iv | Statistics ENTSO-E SO&AF | - | 767.97 | 100.00% |

Table 3 gives an overview of the four objects of comparison and their sizes. Indirectly, it also shows that adding WEPP to the matching increases the cumulative capacity by ~40 GW of the matched dataset, even though this value is still ~20 GW lower than what the statistics report, but already higher ~19 GW higher than the total capacities of the WEPP. While the matched dataset w/ WEPP contains more than twice as much records as the matched dataset w/o WEPP does, it only contains roughly 5.6% more of represented capacity. This is an indication that the set difference represents many comparatively small units. This phenomenon can also be confirmed geographically for the dataset w/o WEPP (cf. Figure 8a) and the one w/ WEPP (Figure 8b) in which the six most important fuel types (the rest summed among 'Other') were plotted onto a European map. The differences can especially be seen for numerous comparatively small gas and hydro plants, visible e.g. in Denmark, France, Portugal and Poland.

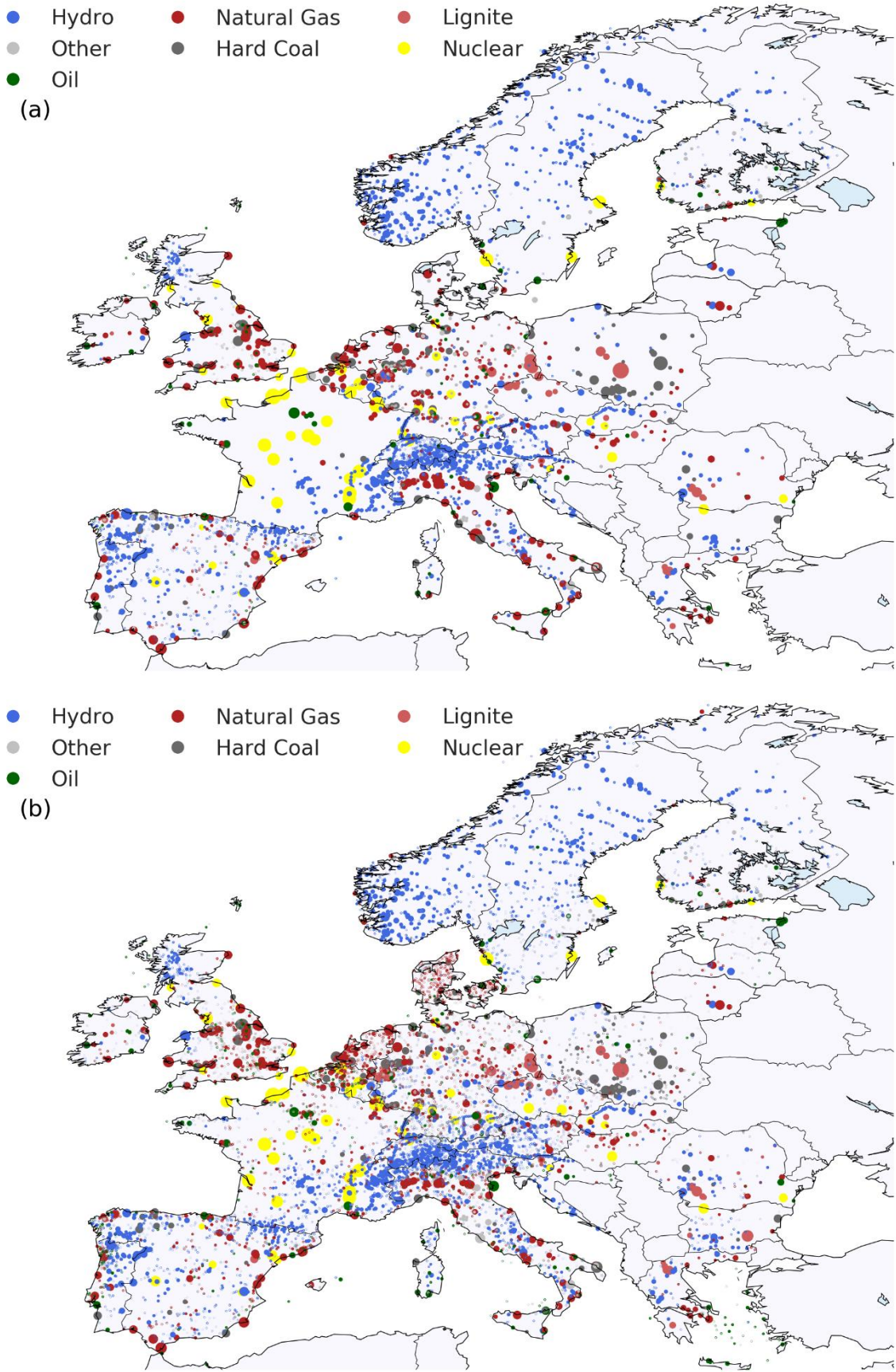

**Figure 8:** Map plots of the matched datasets once without WEPP (a) and once with (b) - showing differences for comparatively small units, visible e.g. in Denmark, France, Sweden and Poland.

Since a bar plot with 28 countries, showing four bars each might easily lead to confusion, we instead decided to plot country-wise (but not fuel type-specific) data points that are shown in Figure 9 in the form of scatter plots for each object of comparison against each another.

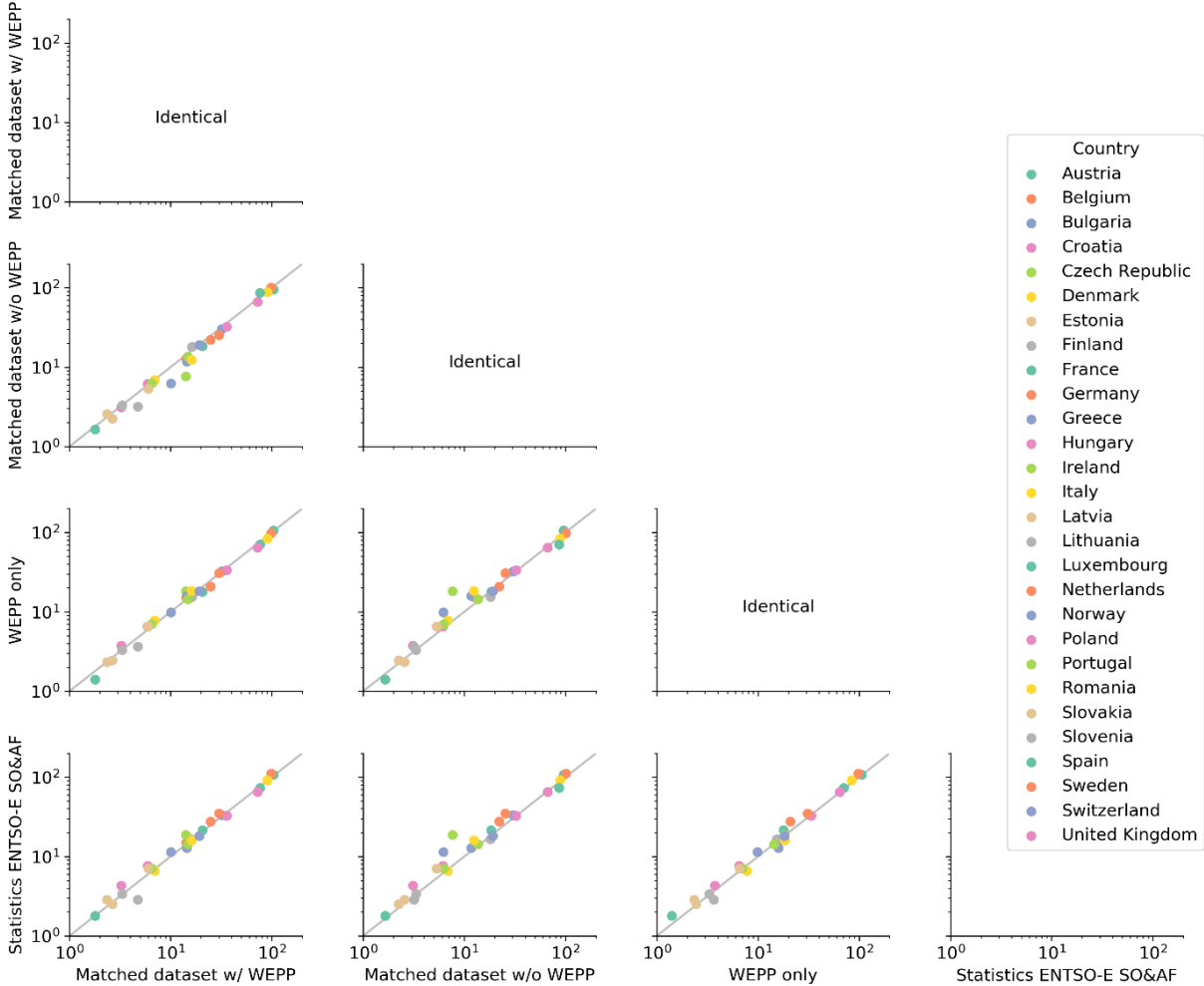

**Figure 9: Scatter plot of installed capacities per country in [GW] for each object of comparison against one another**

The plotted data forms point clouds close to the 45° identity line indicating a very good agreement with $R^2$-values ranging from 98.84% to 99.68%. However, made visible through the double logarithmic axes, all clouds in which the matched dataset w/o WEPP is contained seem to deviate almost entirely to one side, hinting at a slight underestimation of this dataset (e.g. $R^2$=98.84% for matched dataset w/o WEPP vs. statistics), confirming the results from subsection 3.1. Adding the WEPP to the matching contributes to some extent ($R^2$=99.45%) to the matching result when comparing the two subplots on the left in the lowest row. The reason for this is the requirement that a record is only approved during the matching process if it occurs in at least two databases (cf. subsection 2.2). Due to this fact, a lot of records are disregarded in comparison to the matched dataset w/ WEPP in which especially CARMA and WEPP complement each other quite well. Nevertheless, the main finding of these plots is that the total installed capacities for each country are represented to a quite good extend for the three datasets in comparison with the claimed capacity statistics.

Since any fuel type-specific information is hidden in the depiction of Figure 9, we decided to draw country-wise subplots, showing the results for each fuel type on the x-axis and for the capacity on the y-axis, categorized by vertical bars representing the four objects of comparison. Due to the large image size needed to display all the subplots properly, we placed the graph into the appendix (cf. Figure 12). One can see that for most countries and fuel types all four objects of comparison claim capacities in the same order of magnitude. Interestingly, obvious exceptions can be found in a couple of Eastern European countries: In Estonia, the statistics report some 2 GW as hard coal whereas the other three objects report them as oil. This might be explained by the fact that Estonia uses some peat-fired power plants (which can be seen as a form of hard coal or lignite with an even lower "lower heating value"), while peat is often not a fuel-type category in databases. In Lithuania, where the WEPP and matching w/o WEPP claim about 2 GW for natural gas, while the matching w/ WEPP claims more than ~3 GW and statistics around ~2 GW but as "other". In Bulgaria, both the matched dataset w/ WEPP, statistics and WEPP claim ~4 GW of lignite capacity, whereas the matched dataset w/o WEPP claims no capacity at all. Importantly, it needs to be reconsidered that the addition of WEPP to the matching process does not necessarily lead to higher capacity claims. This is due to the fact that the final capacity claim is based on the median of the capacities of the databases with the highest reliability score (cf. subsection 2.2), thus adding a database with a lower capacity record can indeed reduce the capacity claim of this record in the matched dataset. In Germany, the statistics deviate noticeably from the three other objects for natural gas plants. This is due to the fact that there exists a high number of very small 'must-run' CHP units which receive special funding through Germany's CHP law [26], but are not contained in any of the input databases, similarly like wind and solar power. Satisfactorily, in both matched datasets nuclear capacities are matched to a very good extent in all countries, in some even exactly like in Finland, France or Hungary. The same tendency holds true for both hard coal and lignite units, although to a lower extent due to outliers in countries such as Bulgaria, Romania or the Czech Republic. For the fuel-types bioenergy, waste and 'other' no clear tendency can be formulated, since they deviate strongly from country to country, in low orders of magnitude though. These deviations are primarily caused by diverging definitions of those fuel types (e.g. organic waste can be included in either bioenergy or waste[7]).

Since power plants with high capacities are more prominent, they are also likely to be contained in more databases. Therefore we evaluated qualitatively, whether there is connection between the capacity and the number of datasets involved in the matching process for each record of the matched dataset. The results are shown in form of two subplots, both containing scatter plots showing point clouds with one point for each record in Figure 10. The upper subplot (a) distinguishes the matched dataset with WEPP by the number of datasets involved in the match. We find a slight tendency which indicates the assumed relationship that bigger plants are more likely to be matched: The more number of databases are involved in the match, the more does the cloud move towards higher capacities. In general, this trend is also valid when the datasets are filtered by fuel types, therefore we chose to display here independently of fuel types. For a visual comparison, the lower subplot (b) shows clouds for each of the originating datasets on its own, but neither colored nor rescaled in contrast to those in Figure 1.

---

[7] The allocation of fuel-types is being done differently for each database, due to different names and abbreviations. They can be checked directly within the data module of PPM.

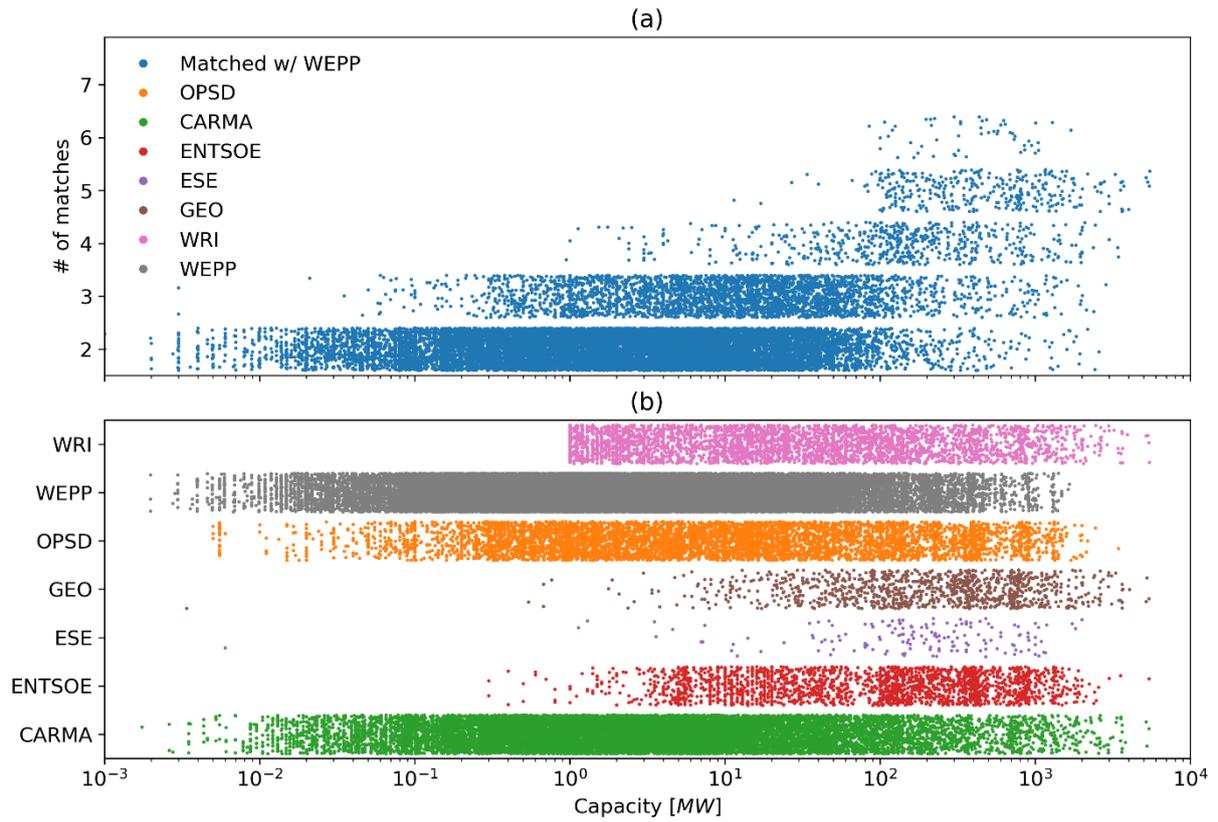

**Figure 10: Scatter plots of capacities by number of datasets involved in match (a) and capacities for each dataset on its own in (b).**

Another aspect under consideration is the variable "YearCommissioned" which is defined as the year in which a power plant had been synchronized with the power grid. Unfortunately, only three out of the seven input databases, namely ESE, OPSD and WEPP, contain data about the commissioning year. Therefore it is very important to note that every positive match, which has been found without participation of one of those databases, does not have a commissioning year entry in the record of the final dataset. Consequently, only those units which have an entry can be depicted in Figure 11, showing four subsets of the development of capacity additions throughout the current and last century. For the two plots at the left, one can see clearly the installation peak of nuclear plants in the 1970s and 1980s and the peak of natural gas installations (or retro-fittings) in the last 25 years. Since wind and solar power have been excluded from the analyses here (cf. subsection 2.4), their enormous capacity additions during the recent years are, of course, not depicted in any of the subplots.

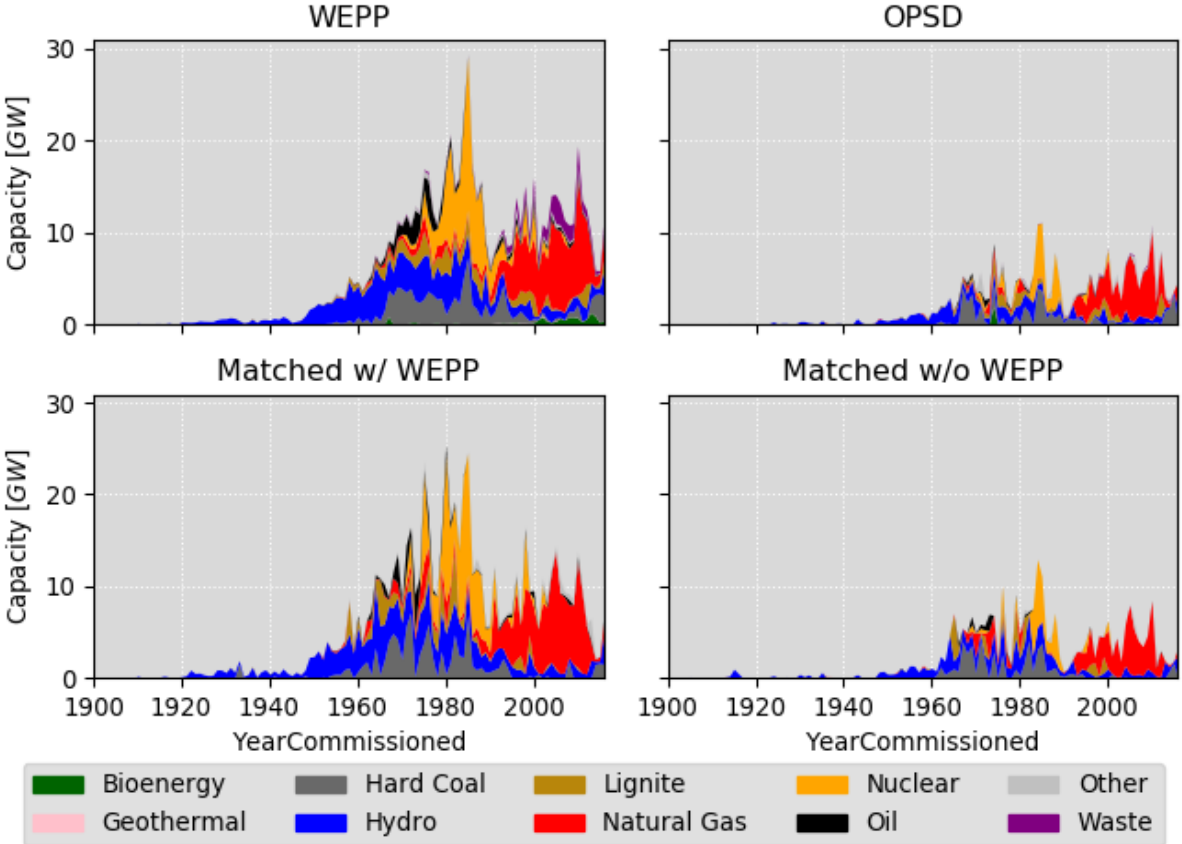

**Figure 11: Subset of development of capacity additions throughout the current and last century**

Apparently, the two plots on the right side (OPSD and matched w/o WEPP) show substantially less capacity additions than the two graphs on the left side. For the matched w/o WEPP dataset 47.36% of the entries and 55.82% in terms of capacity have no data for the commissioning year. However, for the matched w/ WEPP dataset only 10.68% of the entries and only 6.33% in terms of capacity lack data about the commissioning year. The WEPP itself only consists of 1.07% of capacity with no data about the commissioning year, whereas the OPSD misses 46.6% in terms of capacity. Table 4 lists per country the amount of records containing a commissioning year, the absolute amount of records and their ratio both for the matched dataset w/ WEPP and w/o WEPP. A comparison of the ratio columns shows that for almost every country, the WEPP increases the share of records containing commissioning year information. The two exceptions are Germany and Switzerland and it is important to note that only their ratio decreases, whereas the total amount of records with commissioning year increased in both countries.

**Table 4: Comparison of records containing a commissioning year in relation to the count of the two datasets**

|  | Matched dataset with WEPP | | | Matched dataset without WEPP | | |
|---|---|---|---|---|---|---|
|  | Records w/ Year | Total | Ratio | Records w/ Year | Total | Ratio |
| Austria | 643 | 679 | 95% | 151 | 164 | 92% |
| Belgium | 228 | 272 | 84% | 2 | 27 | 7% |
| Bulgaria | 97 | 111 | 87% | 2 | 20 | 10% |
| Croatia | 51 | 51 | 100% | 3 | 24 | 13% |
| Czech Republic | 153 | 178 | 86% | 0 | 19 | 0% |
| Denmark | 416 | 430 | 97% | 14 | 22 | 64% |
| Estonia | 20 | 38 | 53% | 0 | 3 | 0% |
| Finland | 243 | 369 | 66% | 15 | 279 | 5% |
| France | 1134 | 1362 | 83% | 36 | 172 | 21% |
| Germany | 1529 | 2038 | 75% | 510 | 529 | 96% |
| Greece | 122 | 135 | 90% | 2 | 30 | 7% |
| Hungary | 77 | 79 | 97% | 0 | 19 | 0% |
| Ireland | 80 | 108 | 74% | 1 | 17 | 6% |
| Italy | 1451 | 1575 | 92% | 19 | 399 | 5% |
| Latvia | 44 | 47 | 94% | 0 | 4 | 0% |
| Lithuania | 20 | 22 | 91% | 1 | 5 | 20% |
| Luxembourg | 81 | 88 | 92% | 1 | 2 | 50% |
| Netherlands | 573 | 589 | 97% | 3 | 50 | 6% |
| Norway | 784 | 803 | 98% | 4 | 408 | 1% |
| Poland | 397 | 448 | 89% | 6 | 66 | 9% |
| Portugal | 370 | 376 | 98% | 103 | 122 | 84% |
| Romania | 346 | 350 | 99% | 0 | 28 | 0% |
| Slovakia | 41 | 41 | 100% | 36 | 37 | 97% |
| Slovenia | 85 | 89 | 96% | 44 | 54 | 81% |
| Spain | 1417 | 1488 | 95% | 1478 | 2520 | 59% |
| Sweden | 637 | 665 | 96% | 0 | 157 | 0% |
| Switzerland | 662 | 695 | 95% | 522 | 526 | 99% |
| United Kingdom | 1115 | 1222 | 91% | 213 | 311 | 68% |
| **Total** | **12816** | **14348** |  | **3166** | **6014** |  |

Therefore, it is important to note that for modelling exercises that require installation years (e.g. capacity expansion models like energy and/or power system models), a database which covers commissioning years to a major extent like WEPP seems to be strongly needed. Of course, it would be most desirable if open datasets provided more data about commissioning years in future releases.

## 4 Conclusions and Outlook

The aim of this work was to assess the current state of data quality of open conventional power plant databases for energy modelling exercises. The matched dataset w/o WEPP, which is purely based upon free data accounts for ~92% to ~97% of the overall generation capacity in Europe relative to generation capacity statistics from ENTSO-E SO&AF and power plant capacities from WEPP, whereas the matched dataset w/ WEPP likewise accounts for ~97% to even ~103%. The non-represented power plants in between those two matched dataset are often units with small capacities. If the commissioning years of units are required for modelling needs in the final dataset, the WEPP plays a crucial role by filling these data gaps. Therefore, the integration of the proprietary WEPP into the matching process extends the data basis to a certain extent under the matching criteria, which require that a power plant must be confirmed by at least two sources.

One of the main findings is that the algorithmic combination of freely available data sources is not yet on par with the proprietary WEPP database and a significant amount of manual work with attention to detail remains unavoidable; nevertheless the work load has reduced considerably. Of course, it would be most desirable if the final dataset matched perfectly with given capacity statistics, which in turn reflected real-world installed capacities.

As many energy system modelling groups do not have access to the WEPP due to its relatively high costs or cannot choose to use WEPP due to its restrictive license that impedes providing all input data, they rely solely on open databases. One of our next steps will be to evaluate whether and how the matched dataset w/o WEPP can be extended by non-matched units of one or a combination of some the open databases, replicating similar results as the matching including WEPP does. Moreover, it would be favourable to add more plant parameters (e.g. efficiency) in order to contribute to high data quality in open data and facilitate transparency and reproducibility in energy system modelling.

## 5 Contributions

F. Gotzens conceived the idea for this comparison paper, adapted the python code for this analysis, produced all figures and wrote most sections of the manuscript. H. U. Heinrichs wrote parts of the manuscript and contributed to the analysis of the results obtained and supervised the research. J. Hörsch and F. Hofmann had the initial idea for PPM, wrote the majority of the python code and complemented the manuscript. All authors contributed ideas, gave feedback, derived conclusions and helped to improve the manuscript.

## 6 Acknowledgements

The authors would like to thank the German government's "Stromnetze" research initiative for funding this work within the joint research project "CoNDyNet" (03SF0472B).

**Appendix**

Highly detailed results both per country and per fuel-type for each of the four distinguished cases are shown in Figure 12.

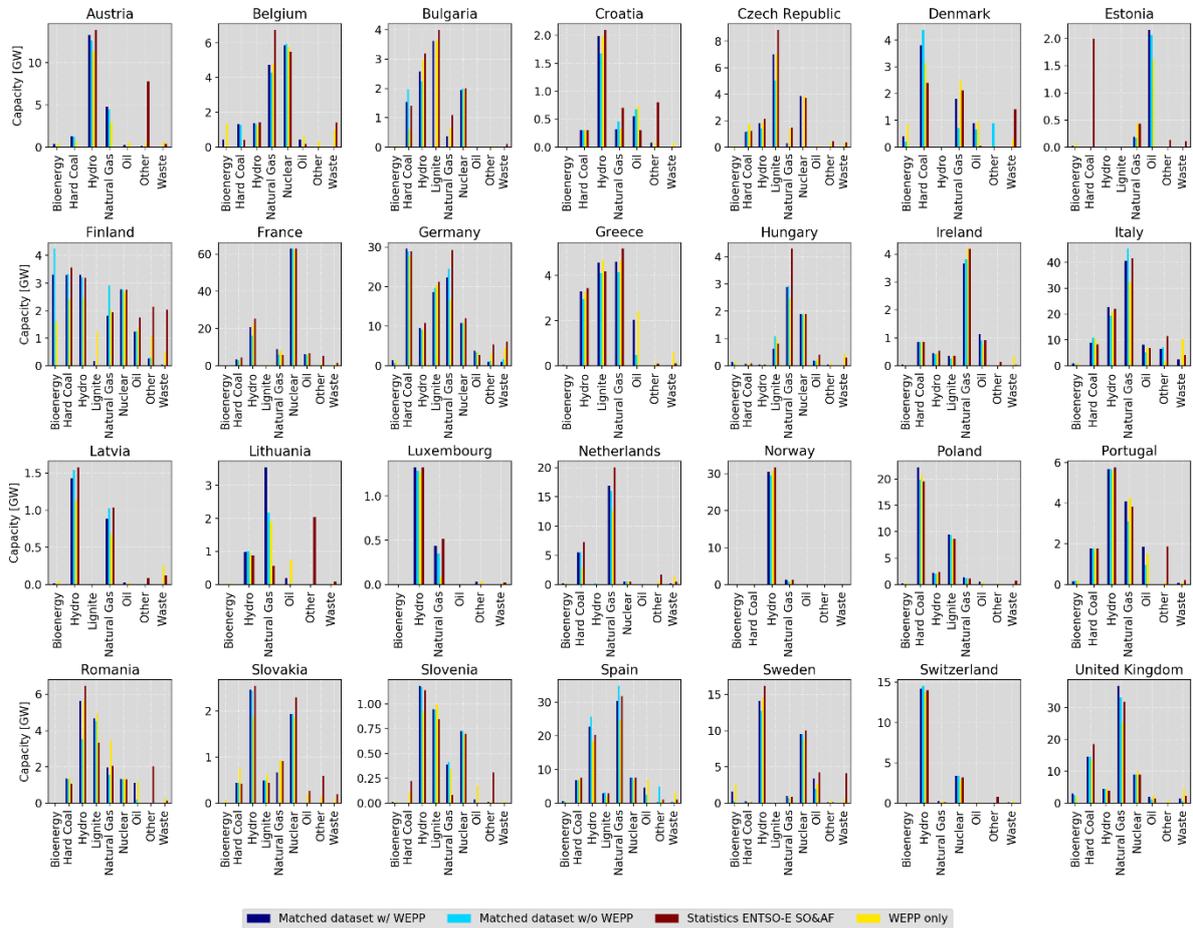

**Figure 12: Comparison results by the two dimensions 'country' and 'fuel type'.**

**Table 5: Reliability scores for each database - higher values indicate higher reliability.**

| Database | Reliability score |
|---|---|
| **BNETZA** | 3 |
| **CARMA** | 1 |
| **ENTSOE** | 4 |
| **ESE** | 4 |
| **GEO** | 3 |
| **OPSD** | 5 |
| **WEPP** | 4 |
| **WRI** | 2 |
| **UBA** | 2 |